\documentclass[twocolumn,showpacs,preprintnumbers,amsmath,amssymb,aps,prb]{revtex4}
\usepackage{graphicx}

\begin{document}
\title{Conductance of 1D quantum wires with anomalous electron-wavefunction localization}
\author{Ilias Amanatidis}
\author{Ioannis Kleftogiannis}
\affiliation{Department of Physics, University of Ioannina, Ioannina 45110, Greece}
\author{Fernando Falceto }
\author{V\'ictor  A. Gopar }
\affiliation{Departamento de F\'isica Te\'orica and Instituto de Biocomputaci\'on y F\'isica de Sistemas Complejos, Universidad de Zaragoza, Pedro Cerbuna 12, E-50009, Zaragoza, Spain.}

\begin{abstract}
We study the statistics of the conductance $g$ through one-dimensional disordered systems where electron wavefunctions decay spatially as $| \psi | \sim \exp (-\lambda r^{\alpha})$ for $0 < \alpha <1$ , $ë$ being a constant. In contrast to the conventional Anderson localization where  $| \psi | \sim \exp (-\lambda r)$ and  the conductance statistics is determined by a single parameter: the mean free path, here we show  that  when the wave function is anomalously  localized ($\alpha <1$) the full statistics of the conductance is determined by the average $\langle \ln g \rangle$ and the power $\alpha$. Our theoretical predictions are verified numerically by using a random hopping tight-binding model  at zero energy, where  due to the presence of chiral symmetry in the lattice there exists
anomalous localization; this case corresponds to the particular value  $\alpha =1/2$.  To test  our theory for other values of $\alpha$, we introduce a statistical model for the random hopping in the tight binding Hamiltonian.

\end{abstract}
\maketitle

\section{introduction}
The phenomena of electron wavefunction localization--Anderson localization--in a disordered media has brought the attention of physicists for several decades. \cite{anderson_0,anderson,lee}
Nowadays signatures of localization have been found in different physical systems. For instance, experiments with light, acoustic waves, microwaves, and cold atoms have reported evidence of  localization. \cite{phystoday,shapiro}

In the standard Anderson localization problem, electron wave functions are localized exponentially in space:
\begin{equation}
| \psi|  \sim \exp{(-\lambda r)} ,
\end{equation}
where $\lambda$ can be identified as the inverse of the localization length. For practical purposes, it is more convenient to define the localization length through measurable transport quantities; for a system of length $L$, the localization length is defined by the exponential decay of  the dimensionless conductance $g$, or transmission. Since $g \propto |\psi(L)/\psi(0)|^2$ we have
that
$
 g \propto \exp{(-2 \lambda L) }.
$
Thus,   the inverse localization length $\lambda$ is usually estimated by the relation
\begin{equation}
\label{lng}
\langle - \ln g \rangle = 2\lambda L  ,
\end{equation}
i.e.,  the average  $\langle  \ln g \rangle$ is a linear function of $L$ in the standard electron localization problem.
Within a non-interacting electron model,  a scaling approach of localization has  successfully described the statistical properties of  electronic transport. \cite{melnikov,dorokhov,mello_groups, mello_book} Within this approach, it has been found that the complete distribution of the dimensionless conductance is determined by a single parameter: the inverse localization length\cite{cohen}, given by Eq. (\ref{lng}). In general,  one might say that there is a good understanding of the statistical properties of the transport in the Anderson localization problem in one dimensional (1D) and quasi-one dimensional disordered systems.

On the other hand, anomalous localization of electron wave functions  has been found in 1D disordered systems, \cite{soukoulis,ziman,inui,spiros1}  against the general idea that in 1D systems all the electronic eigenstates are always exponentially localized.
This problem has been much less studied than the above standard localization phenomena. For instance, a disordered system described by a random hopping tight binding  model was studied in Ref. \onlinecite{soukoulis}, where it was found that the typical conductance ($\exp{\langle \ln g \rangle}$)  behaves as
\begin{equation}
g_{\mathrm{typ}}  \propto \exp(-\lambda {L}^{1/2}) .
\end{equation}
This unconventional localization of electrons (also named delocalization\cite{spiros1}) can be explained  by the presence of a symmetry in the lattice, the so-called chiral symmetry, \cite{inui,spiros1} which makes the energy spectrum symmetric around  zero energy. \cite{soukoulis}
The effects of the chiral symmetry in a disordered system was studied also within a scaling approach to localization. \cite{brouwer1, mudry_brouwer_furusaki} It was  found  that there is no exponential localization of the conductance  and the  logarithm of $g$ is not self-averaging, while the ensemble average
$\langle \ln g\rangle$ is not proportional to $L$, as in the standard Anderson localization, but to ${L}^{1/2}$, i.e., $\langle \ln g\rangle \propto {L}^{1/2}$.  A similar delocalization has been found in disordered superconducting wires,  \cite{mudry_brouwer_furusaki,brouwer_furusaki_gruzberg_mudry,motrunich,brouwer_furusaki_mudry,gruzber}  where
the Bogoliubov-de Gennes Hamiltonian has additional symmetries. \cite{evers_mirlin}
Delocalization at zero energy has been also studied using tight-binding models of spinless fermions with particle-hole symmetric disorder \cite{balents} and in 1D systems in the context of phase transitions in random XY spin chains, \cite{ross}  which is mapped onto the so-called random mass Dirac model; within this model, it was also found \cite{steiner_1, steiner}  that $\langle -\ln g\rangle \propto {L}^{1/2}$. In addition, statistical properties of the conductance in 2D systems under the presence of chiral symmetry has been studied in Ref.  \onlinecite{verges}.

In the present paper we show that the complete distribution of the conductance for anomalous transport (nonstandard exponential localization)  can be  determined  by the value of the average $\langle \ln g \rangle$ and the power  $\alpha$  of its dependence on length $L$, i.e., $\langle \ln g \rangle \propto L^\alpha$.  Thus, within a model of noninteracting electrons,  the microscopic details of the systems (Hamiltonian) do not enter into the description of the statistical properties of the transport, in this sense,  the description is \emph{ universal}. Our theoretical model is based on a previous study of the conductance statistics  of  1D disordered quantum wires where the random configuration of  potential scatterers along the wire follows a distribution with a long tail (Levy-type distribution). \cite{fernando_gopar} However, in that paper,
the analysis of the transport was restricted to disordered wires where information on the L\'evy-type distribution was explicitly introduced  into the disorder configuration of the scatterers.
Here, we do not need a Levy-type disorder configuration but a mechanism to produce anomalous localization of the electron wave function, within a single-electron model, e.g. the chiral symmetry.  Thus,  as we show in this work,  the results in Ref. \onlinecite{fernando_gopar} can be applied in general to disordered systems where electron anomalous localization is present. This larger scope of such  statistical analysis was overlooked in Ref. \onlinecite{fernando_gopar}.

The remainder of this paper is as follows, after presenting a brief review of the results for wires with L\'evy-type disorder, we introduce the random hopping tight binding model where at zero energy  anomalous localization is present. The numerical results of this model will be compared with our theoretical predictions; in particular, we are interested in  the conductance distribution. The numerical results from the random hopping tight binding model at zero energy  corresponds to a special case of our theory ($\alpha=1/2$). To go further and verify our results in a more general way,  we introduce a statistical model for the random hopping which allows to study different degrees of localization characterized by the value of $\alpha$. We finally summarize our results and give some conclusions in  the last part of the paper.
\section{Theoretical Model}
As we have mentioned, our theoretical model of this work is based on a study of coherent transport in the presence of L\'evy-type disorder. \cite{fernando_gopar}  We briefly mention
that L\'evy-type random processes are described by a density probability
$q_{\alpha,c}(x)$ with a long tail: for large $x$, $q_{\alpha,c}(x) \sim c/x^{1+\alpha}$ with $0 < \alpha <2$ and $c$ being a constant. These kind of distributions are also known by mathematicians as $\alpha$-stable distributions. \cite{levy-kolmorogov-calvo,Gnedenko,Calvo,uchaikin} Notice that  first and second moments diverge for
$0< \alpha <1$.  Motivated by  the realization of experimentally controlled L\'evy processes in the so-called L\'evy glasses, \cite{barthelemy} in Ref. [\onlinecite{fernando_gopar}]  a  model was developed  to describe the statistical properties of the conductance through a 1D quantum wire where electrons suffer multiple scattering  due to  scatterers  placed along the wire in a random way accordingly to a  L\'evy-type distribution (see [\onlinecite{mercadier,boose,raffaella,beenakker,fernadez}] for other examples  where L\'evy processes have been studied in connection  to transport problems). It was found in [\onlinecite{fernando_gopar}] that  the full statistics of the conductance is determined by the average $\langle \ln g \rangle$ and the exponent $\alpha$ of power-law tail in the macroscopic limit ($L \gg  c^{1/\alpha}$).  In particular,  it was shown that  the complete distribution of conductances $P_\xi(g)$, with $\xi =\langle \ln g \rangle$, is given by
\begin{eqnarray}
\label{pofG_xi}
P_\xi(g)=\int_0^\infty p_{s(\alpha,\xi,z)}(g) q_{\alpha,1}(z){\rm d}z ,
\end{eqnarray}
for $\alpha <1$, where $q_{\alpha,c}$ is the probability density function of the L\'evy-type distribution supported in the positive semiaxis,
$
s(\alpha,\xi,z)={\xi}/(2{z^\alpha I_\alpha)},
$
$
I_\alpha =1/2 \int_{0}^{\infty} z^{-\alpha} q_{\alpha,1}dz
$, and
\begin{equation}
\label{pofg}
p_s(g)=\frac{s^{-\frac{3}{2}}}{\sqrt{2\pi}}
\frac{{\rm e}^{-\frac{s}{4}}}{g^2}\int_{y_0}^{\infty}dy\frac{y{\rm e}^{-\frac{y^2}{4s}}}
{\sqrt{\cosh{y}+1-2/g}},
\end{equation}
where  $y_0={\rm arcosh}{(2/g-1)}$. Also, it was shown that the average of the logarithm of the conductance  depends on $L$ as
\begin{equation}
\label{lngalpha}
 \langle- \ln g \rangle \propto L ^\alpha ,
\end{equation}
for $0 < \alpha < 1$,  while for values  $1 \le \alpha < 2$ the linear behavior ($ \langle- \ln g \rangle \propto L $)  is recovered. From the same model one  can also find that conductance average behaves as
\begin{equation}
\label{averg}
 \langle g \rangle \propto L^{-\alpha} ,
\end{equation}
for $0< \alpha < 1$,  in contrast to the exponentially dependence with $L$  in the standard  localized regime.
The most interesting effects of anomalous localization are seen for values $0< \alpha < 1$, so we concentrate in this region, although the case
$1 \le \alpha < 2$ can be analyzed within the same theoretical framework.

\begin{figure}
          \centering
          \includegraphics[width=0.8\columnwidth]{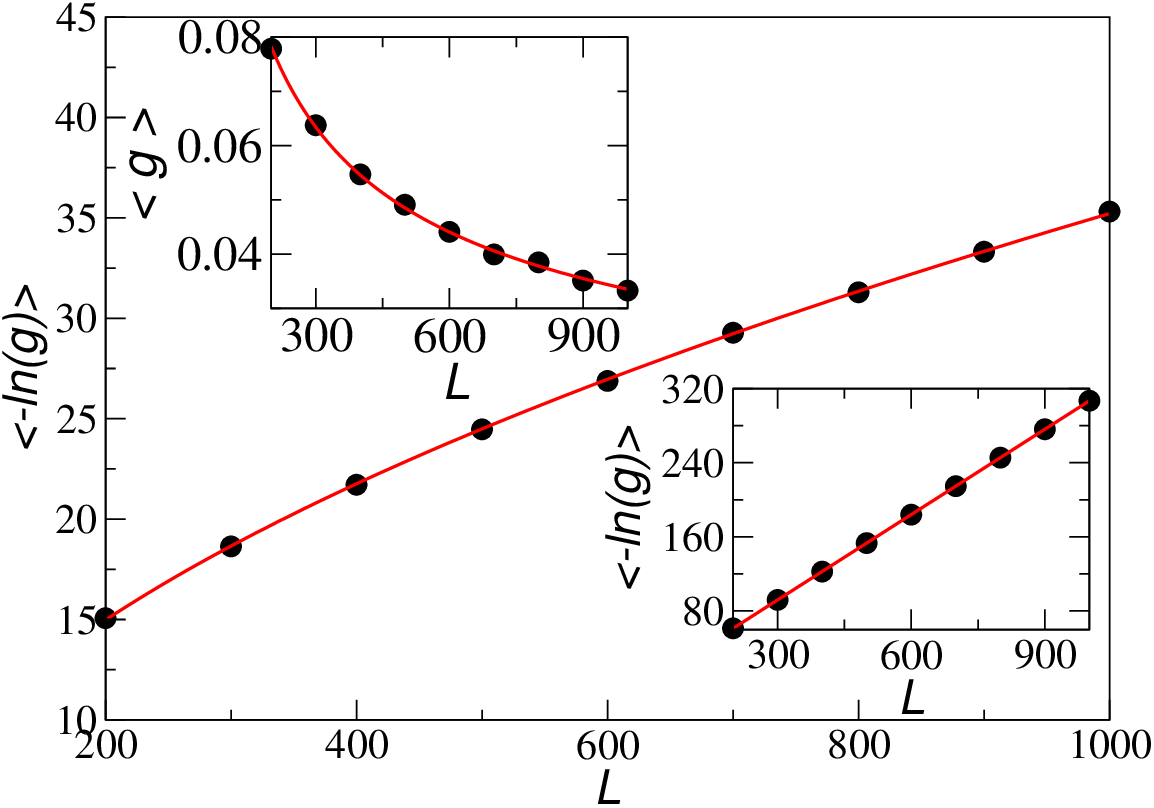}
          \caption{$\langle - \ln g \rangle$ as a function of the length $L$ at energy $E=0$ for strength of disorder $w=2.5$. The solid line is obtained by fitting the data (dots) according to Eq. (\ref{lngalpha}) with $\alpha=1/2$. Upper inset:$\langle g \rangle$ as a function of the length $L$ (for the same parameters in the main frame). The solid line is fitted to the numerical data assuming that $\langle g \rangle \propto L^{-1/2}$.
          A good agreement is seen.
          Lower inset: $\langle - \ln g \rangle$  for a linear chain with off-diagonal logarithmic disorder as a function  $L$ at energy $E=0.1$  and strength of disorder $w=2.5$  (50000 realizations). As expected,  a linear behavior is observed indicating Anderson localization.}
          \label{Fig1}
\end{figure}
\section{anomalous localization: $\alpha=1/2$}
Next we consider the  tight binding model with nearest neighbor  random hopping,  at zero energy, described by the Hamiltonian
\begin{equation}
\label{tight}
H= \sum_n t_n ( c_{n}^\dagger c_{n+1} +c_{n+1}^\dagger c_{n} ) ,
\end{equation}
where $c_n^\dagger$ and $c_n$ are creation and annihilation operators for spinless fermions, and $t_n (>0)$ are the random hopping  elements sampled from a distribution of the  form $P(t)=1/wt,
\exp(-w/2) \leq t \leq \exp (w/2)$, where $w$ denotes the strength of the disorder. This is the so-called logarithmic off-diagonal disorder. \cite{soukoulis} As we have mentioned, the model described by Eq. (\ref{tight}) has been found to present  unconventional localized states at zero energy,  \cite{soukoulis,ziman,inui,spiros1,steiner,steiner_1,balents} whereas for nonzero energy standard localized states are present. To illustrate this fact,  we have calculated the conductance within the Landauer-B\"uttiker approach. In Fig. \ref{Fig1}  we show the ensemble average $\langle \ln g \rangle$ as a function of the length of the system (in units of the lattice constant) at zero and nonzero energies. As we can observe $\langle -\ln g \rangle \propto L^{1/2}$ at zero energy (main frame), while a linear dependence on $L$  is obtained at finite energy (lower inset), restoring the standard  Anderson localization. Additionally,  in the upper inset of  Fig. \ref{Fig1} we show the average of the conductance $\langle g \rangle$ at zero energy, which depends on the length as $L^{-1/2}$, as given by Eq. (\ref{averg}).

\begin{figure}
          \centering
           \includegraphics[width=0.75\columnwidth,angle=0]{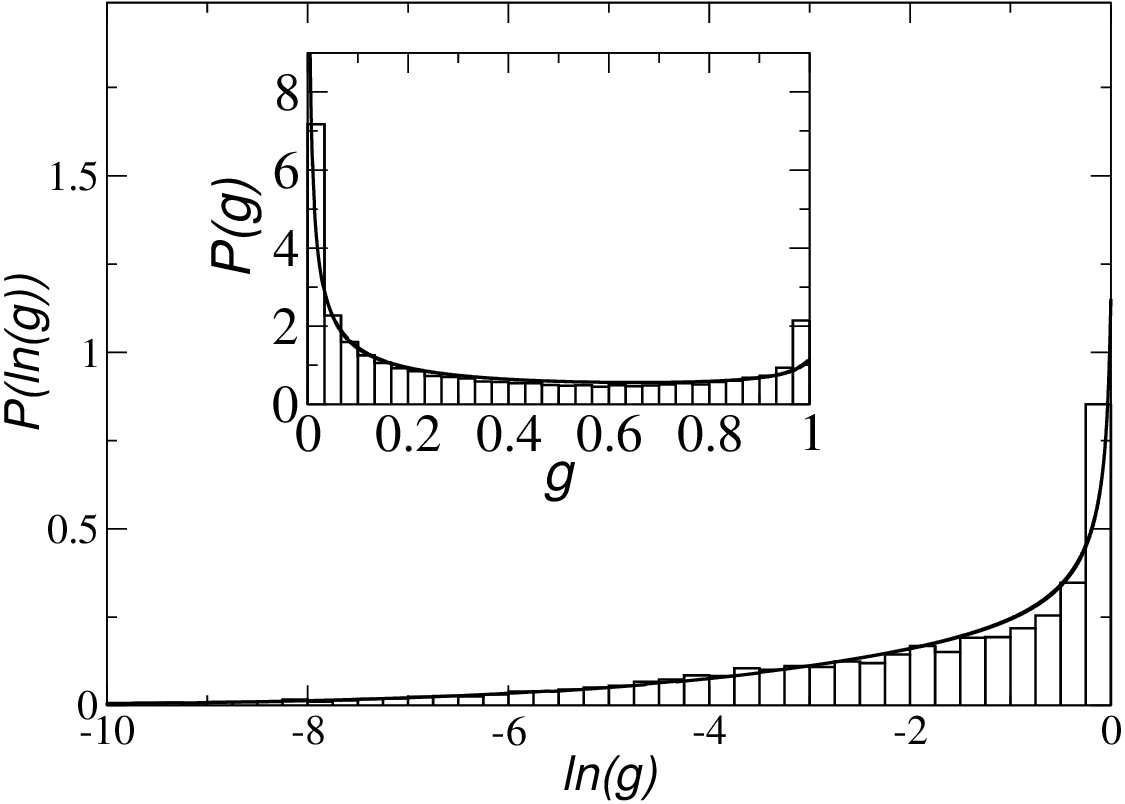}
          \caption{The distribution of $\ln g$ for a system of length $L=400$  with offdiagonal logarithmic disorder, at energy $E=0$ and strength of disorder $w=0.35$ (50000 realizations). From the numerical data $\langle -\ln g \rangle = 2.1$. Using this information the theoretical distribution (solid line) is calculated with $\alpha=1/2$, Eqs. (\ref{pofG_xi}) and (\ref{pofg}). Inset: $P(g) $ for the same case as in the main frame. The coexistence of insulating and ballistic regimes are manifested by the presence of two peaks
          at $g=0$ and 1. As we can see, the theory (solid line) gives correctly the trend of the numerical results (histograms).
          }
          \label{Fig2}
\end{figure}

\begin{figure}
          \centering
         \includegraphics[width=0.75\columnwidth]{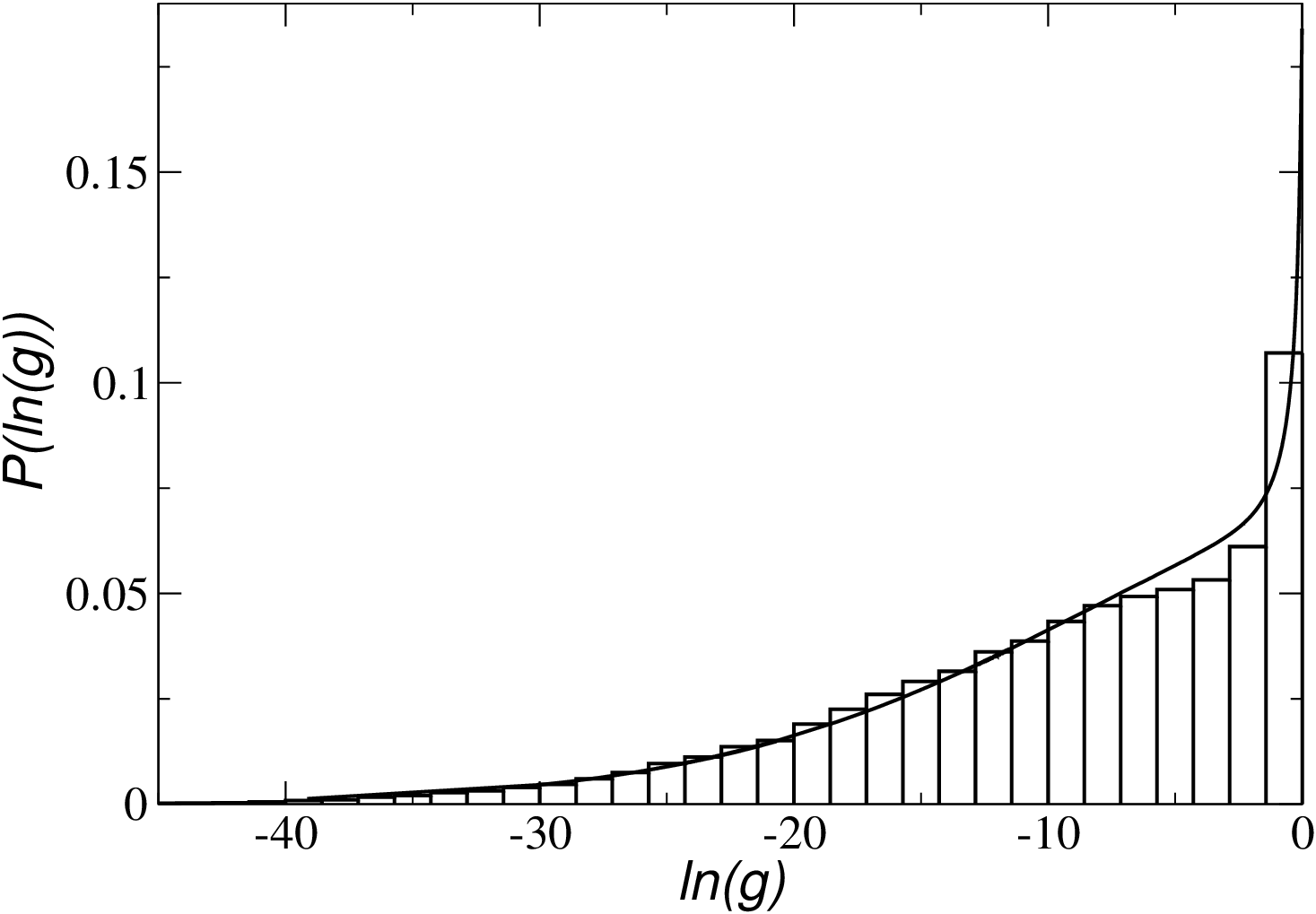}
          \caption{ The distribution of $\ln g $ for strength of disorder $w=1.2$. $\langle -\ln g \rangle =9.7 $ corresponding to a more insulating case than the previous one (Fig. \ref{Fig2}), while the power-law dependence on $L$ remains $1/2$ (Fig. \ref{Fig1}). A good agreement is seen between the numerical histogram and the corresponding theoretical distribution (solid line).}
          \label{Fig3}
\end{figure}

We now show that the complete distribution of conductance is described by Eq. (\ref{pofG_xi}).  As we have claimed, in order to compare the theoretical and numerical results, we only need the
information of the value $\langle \ln g \rangle$ and its power dependence on $L$, which are taken from the numerical simulation; thus, there is no free parameters in our theory.  In Figs. \ref{Fig2}  and \ref{Fig3} we show  the distribution of the conductance obtained from the numerical simulations (histograms) for two different strengths of disorder and the corresponding theoretical distributions (solid lines) accordingly to Eq. (\ref{pofG_xi}). Note that we plot $P(\ln g)$ in the main frames, instead of $P(g)$, since for very insulating cases the details of the distributions are better seen in this way. For the smaller case of strength disorder ($w=0.35$) in Fig. \ref{Fig2} we have included $P(g)$ in a inset. Here we can observe two peaks at $g=0$ and $g=1$,  which is due to the existence of  strong sample-to-sample conductance fluctuations, i.e., in our ensemble a considerable amount of samples behaves like insultors ($g << 1$), whereas another important amount of them behaves as ballistic samples  ($g \approx 1$). This behavior is very robust in the sense that if we increase the length of the system or the disorder degree the peak at $g=1$ survives.
%This means that insulating ($g<<1$) and ballistic regimes ($g \approx 1)$ coexist.
This  is not seen in the conventional 1D  electron localization problem. In Fig. \ref{Fig3} we increase the strength of disorder to  $w=1.2$. Thus,
for both strengths of disorder, Figs. \ref{Fig2} and \ref{Fig3} show that  our  theory gives correctly the trend of the numerical distribution. We might see a small difference between numerics and theory in the inset of Fig. \ref{Fig2} at $g \approx 1$, but we would like to remark that there is not free parameter in our theory.   Therefore our model with $\alpha =1/2$ describes correctly the statistics of the conductance when anomalous localization of the wave function is of the form $|\psi| \sim \exp{(-\lambda {L}^{1/2})} $. However, this is a special case for our model. We would like to explore different exponential power decays $\alpha$ of the wave function.

%%%%%%
%%%%%%%%
\section{anomalous localization: arbitrary $\alpha$}
In order to investigate different anomalous-localization degrees of the wave function,  we  introduce a statistical model for the nearest-neighbor random hopping model, Eq. (\ref{tight}). In fact, what we need is a model that induce large fluctuations of  the conductance.
A way to introduce such  a  large fluctuations is to consider the  hopping $t_n$  as a random variable that  follows a distribution with a long tail, i.e., a L\'evy-type distribution, and  keeping fixed  the total sum of the hopping elements: $T=\sum _n t_n$. By varying the value of $T$ we can  change the degree of the localization of  the disordered samples. We have verified  numerically that $T$ acts similarly to the length $L$ in the Levy-type configurational disorder used in \onlinecite{fernando_gopar}. However, the tight binding model is more appropriate for numerical simulations. The study is carried out at non-zero energies in order to get rid of the effects of chiral symmetry.

\begin{figure}
          \centering
          \includegraphics[width=0.8\columnwidth]{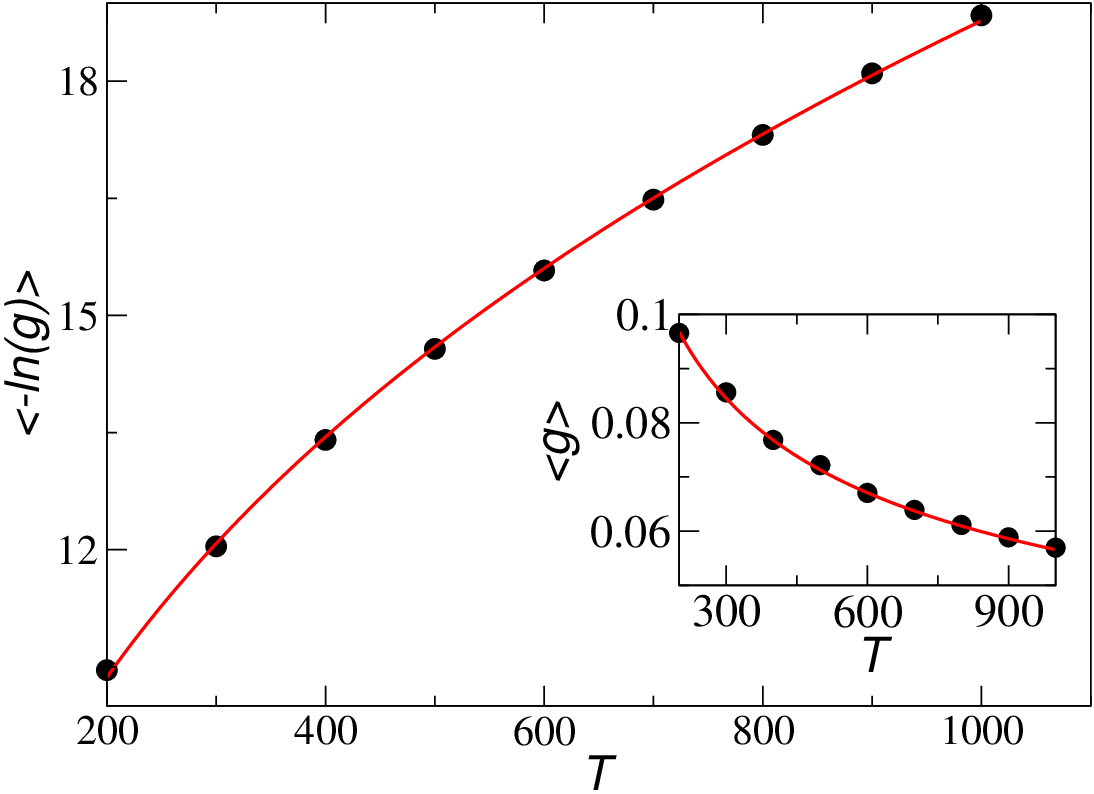}
          \caption{ $\langle -\ln g \rangle$ and $\langle g \rangle$ (inset) as a function of the variable $T$ for  $\alpha=1/3$ in the statistical model  for the tight binding Hamiltonian (see text).  50000 realizations are considered and energy $E=0.1$. The solid lines are  obtained by fitting the power dependence: $T^{1/3}$ and $T^{-1/3}$ for $\langle -\ln g \rangle$ and $\langle g \rangle$, respectively,  as predicted by the theoretical model.}
          \label{Fig4}
\end{figure}

\begin{figure}
          \centering
          \includegraphics[width=0.8\columnwidth]{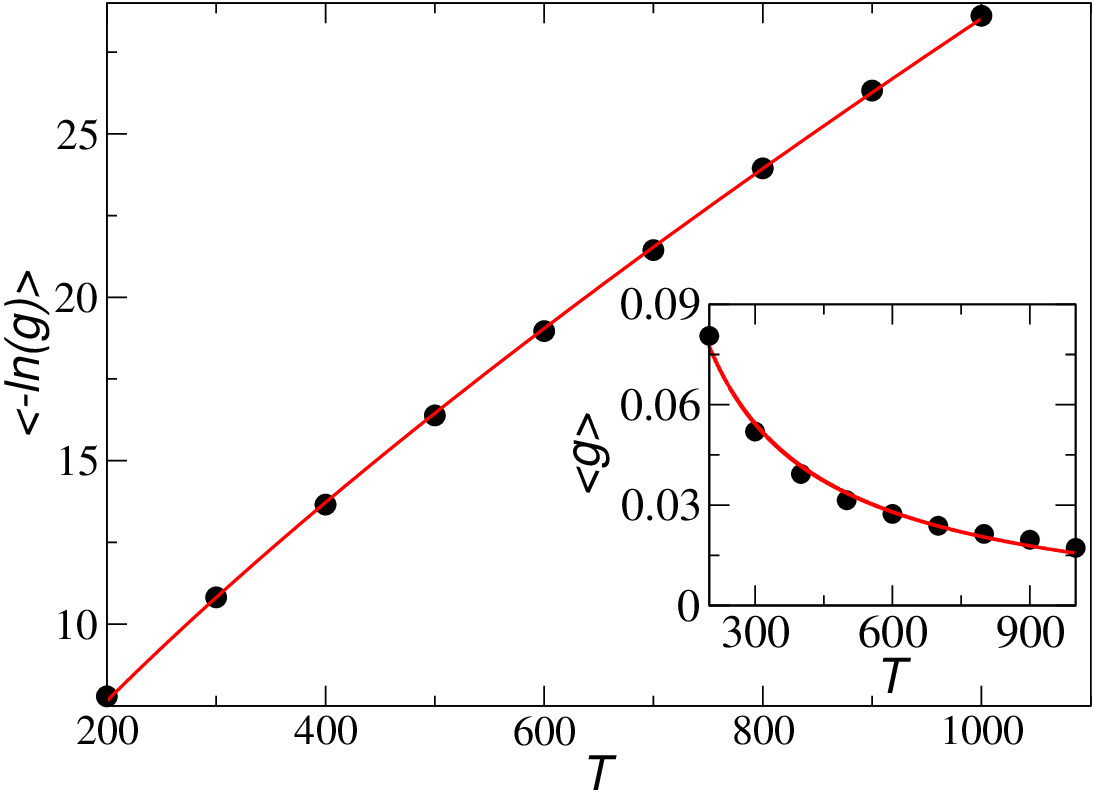}
          \caption{Numerical data (dots) of  $\langle -\ln g \rangle$ and $\langle g \rangle$ (inset) from the tight binding model  at energy $E=0.1$ and $\alpha =3/4$. The solid lines are fitted assuming  that $\langle -\ln g \rangle \propto T^{3/4}$ and $\langle g \rangle \propto T^{-3/4}$, in agreement with  the model, Eqs. (\ref{lngalpha}) and (\ref{averg}).}
          \label{Fig5}
\end{figure}

With the above statistical model for the random hopping tight binding Hamiltonian, we calculate the statistics of the conductance.  The data are collected over an ensemble of $50000$ realizations of disorder. In Figs. \ref{Fig4} and \ref{Fig5} we show first the results for the average $\langle -\ln g \rangle$ and $\langle g \rangle$ (insets) as a function of $T$ where the random hopping elements are generated from  two different  L\'evy-type distributions with tail decay exponents $\alpha=$1/3 and  3/4. We can see that indeed  $\langle -\ln g \rangle \propto T^\alpha$ and $\langle g \rangle \propto T^{-\alpha} $, for both values of $\alpha$.  Having in mind  that $T$ plays  a similar role as $L$ in our configurational disorder model, \cite{fernando_gopar}
we expect that the wave function is anomalously localized as $| \psi | \sim \exp{(-\lambda L^{1/3})} $ and $| \psi | \sim \exp{(-\lambda L^{3/4})}$,  for  $\alpha =1/3$ and $3/4$, respectively.


\begin{figure}
          \centering
           \includegraphics[width=0.8\columnwidth]{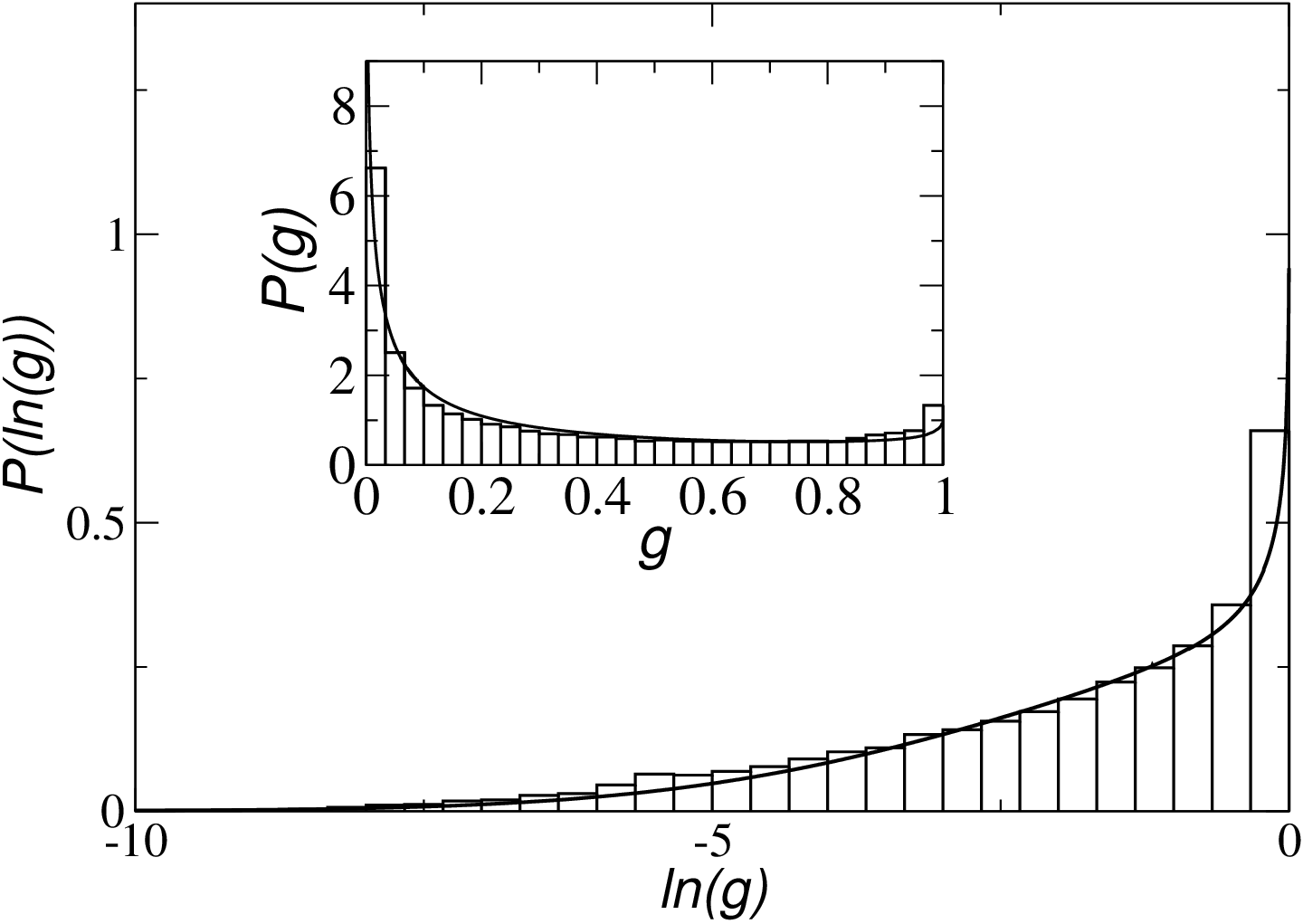}
          \caption{ The numerical distribution  $ P(\ln g) $ (histograms) for $\alpha=3/4$ with  $E=0.1$, $T=35$, and $\langle -\ln g \rangle = 2.0 $ .  Inset: $P(g)$ for the same parameters of  the main frame. The solid line is obtained accordingly to Eq. (\ref{pofG_xi}). A good agreement between numerical and theoretical  (solid line) results is seen. }
          \label{Fig6}
\end{figure}

\begin{figure}
          \centering
           \includegraphics[width=0.8\columnwidth]{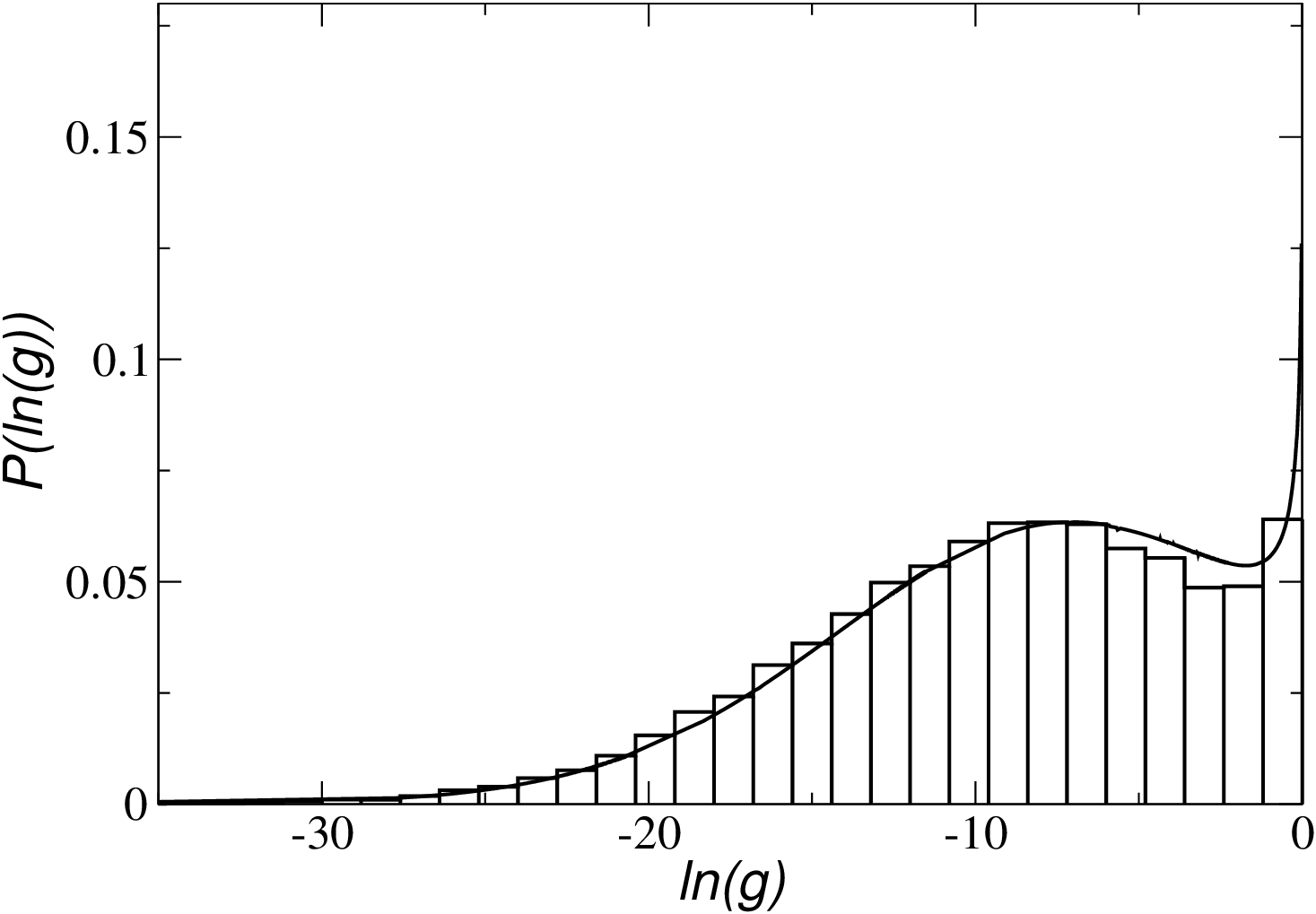}
          \caption{ The distribution of $\ln g $ for $\alpha=3/4$,  $T=250$ and  $E=0.1$. For this case $\langle -\ln g \rangle =9.3 $. We can see that the theoretical result (solid line) describes correctly the numerical distribution.}
          \label{Fig7}
\end{figure}

\begin{figure}
          \centering
           \includegraphics[width=0.75\columnwidth]{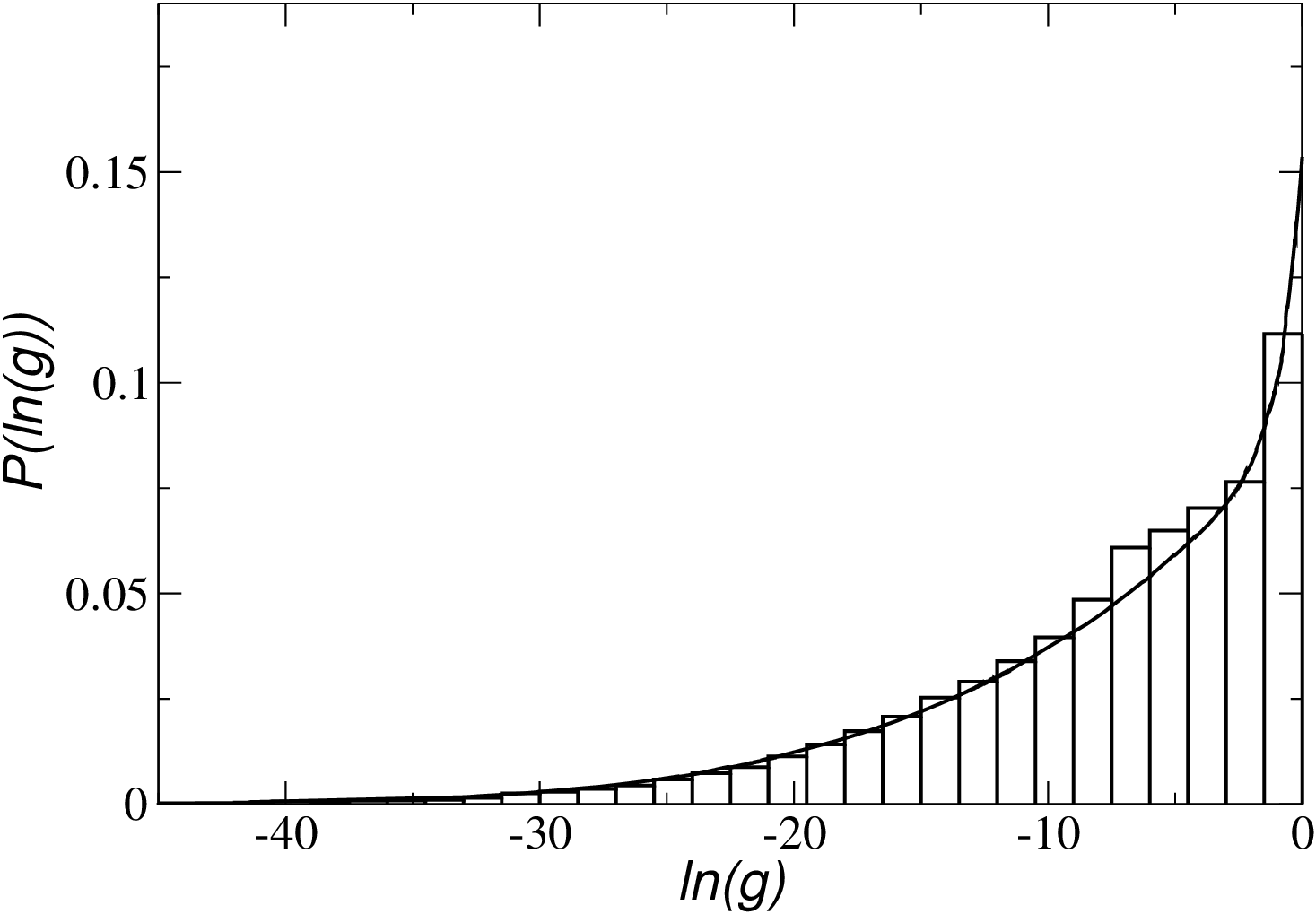}
          \caption{ The distribution of $\ln g $ for $\alpha=1/3$, $T=100$ and energy $E=0.1$. $\langle -\ln g \rangle = 9.4 $ for this case.
           Comparison with the corresponding theoretical distributions is shown. A good agreement between theory and numerics is seen.}
          \label{Fig8}
\end{figure}

We now show that the distribution of the conductance is described by Eq. (\ref{pofG_xi}).  For $\alpha=3/4$ and two different values of
$T$, in Figs. \ref{Fig6} and \ref{Fig7} we compare the numerical simulations (histograms) and the corresponding theoretical results (solid line). The case in Fig. \ref{Fig6} is less insulating than that one  in Fig. \ref{Fig7}, so we plot in an inset  the distribution $P(g)$. For the more insulating case (Fig. \ref{Fig7}), we can observe a nonconventional shape of the distribution $P(\ln g)$.   We mean by nonconventional shape the non Gaussian shape of the  distribution; we recall that  for the standard Anderson localization it is expected a log-normal distribution in the insulating regime.  Thus, from both Figs. \ref{Fig6} and \ref{Fig7}  we can see that the trend of the numerical distributions are well described by our theory. Finally in Fig. \ref{Fig8} we show the distribution $P(\ln g)$ for $\alpha=1/3$. Here we also note the nonconventional shape of the distribution which is a consequence of the anomalously large conductance fluctuations.
%%%%%%%%%
%%%%%%%%%%%%%

\section{conclusions}
To conclude, in this work we have shown that the complete statistics of the conductance of an 1D disordered system,  when electron wave functions are anomalously localized ( $\psi \sim \exp{(-\lambda r^\alpha)}$, $0 < \alpha <1$),  is determined by the exponent $\alpha$  and the average $\langle \ln g \rangle $. In contrast, in the standard Anderson localization, the knowledge of  $\langle \ln g \rangle $ is enough to describe the statistical properties of the conductance. We have verified our results for different values of $\alpha$. For the particular case of $\alpha =1/2$, we have used a random hopping tight binding Hamiltonian at zero energy to verify our predictions since it is well known that nonexponential localization in this model is present due to the existence  of chiral symmetry on the lattice. In order to study  other degrees of anomalous localization (different  values of $\alpha$) we have introduced a statistical  model for the hopping in a tight binding Hamiltonian
that promote the presence of large fluctuations of the conductance. We remark that  our theoretical model do not make any reference to a specific Hamiltonian system  and there is no free parameter; the information needed in our theoretical model ($\alpha$ and $\langle \ln g \rangle $) is extracted from the numerical simulation.
On the other hand,  we have restricted  our study to 1D systems (one channel), we think
an extension to  multichannel systems is of interest since other regimes of transport, e.g. the diffusive regime, can be analyzed. Finally, the conductance statistics in the conventional Anderson localization problem has been extensively studied, we hope this work helps to the understanding of  a much less studied topic in quantum transport: the statistical properties  of the conductance when electron wave functions are anomalously localized.

We acknowledge  support from the MICINN (Spain) under Projects  FIS2009-07277, FPA2009-09638 and DGIID-DGA (Grant No. 2010-E24/2). I. A. thanks the Departamento de F\'isica Te\'orica,  Universidad de Zaragoza, for its hospitality during his visit.

\end{document}